\documentclass[aps,prl,showpacs,twocolumn]{revtex4}
\usepackage{epsfig,amsmath,color,amssymb}
\usepackage[10pt]{moresize}

\begin{document}

\title{A Robust Atom-Photon Entanglement Source for Quantum Repeaters}

\author{Shuai Chen$^{1}$}
\author{Yu-Ao Chen$^{1}$}
\author{Bo Zhao$^{1}$}
\author{Zhen-Sheng Yuan$^{1,2}$}
\author{J\"{o}rg Schmiedmayer$^{3,1}$}
\author{Jian-Wei Pan$^{1,2}$}
\address{$^1$Physikalisches Institut, Ruprecht-Karls-Universit\"{a}t Heidelberg, Philosophenweg 12,
69120 Heidelberg, Germany}
\address{$^2$Hefei National Laboratory for Physical Sciences at Microscale and Department of Modern Physics,
University of Science and Technology of China, Hefei, Anhui 230026, China}
\address{$^3$Atominstitut der \"{o}sterreichischen Universit\"{a}ten, TU-Wien, A-1020 Vienna Austria}
\date{\today}

\begin{abstract}

We demonstrate a novel way to efficiently and very robust create
an entanglement between an atomic and a photonic qubit. A single
laser beam is used to excite one atomic ensemble and two different
spatial modes of scattered Raman fields are collected to generate
the atom-photon entanglement. With the help of build-in quantum
memory, the entanglement still exists after 20.5 $\mu$s storage
time which is further proved by the violation of CHSH type Bell's
inequality. Our entanglement procedure is the building block for a
novel robust quantum repeater architecture [Zhao \emph{et~al},
Phys. Rev. Lett. \textbf{98}, 240502 (2007)]. Our approach can be
easily extended to generate high dimensional atom-photon
entanglements.

\end{abstract}



\maketitle

Quantum communication provides an absolutly secure approach to
transfer information by means of quantum cryptography or faithful
teleportation of unknown quantum states. Unfortunately, the photon
transmission loss and the decoherence scale exponentially with the
length of the communication channel. This makes it extremely hard
to deliver quantum information over long distance effectively. A
quantum repeater protocol
\cite{BriegelPRL1998} 
combining the entanglement swapping, purification and quantum
memory provides a remarkable way to establish high-quality
long-distance quantum networks, and makes the communicating
resources increase only polynomially with transmission distance.

Following a scheme proposed by Duan, Lukin, Cirac, and Zoller
(DLCZ)\cite{DuanNature2001}, in recent years, significant
experimental advances have been achieved towards the
implementation of the quantum repeater protocol by using the
atomic ensemble and linear optics [3-9]. However, the DLCZ
protocol has an inherent drawback which is severe enough to make a
long distance quantum communication extremely difficult
\cite{zhaoboPRL2007,ChenZBquantph2006}. The phase fluctuation
caused by path length instability over long distance is very hard
to overcome. Recently, a more robust quantum repeater architecture
was proposed to bypass the phase fluctuation over long distance
\cite{zhaoboPRL2007,ChenZBquantph2006}. This architecture is based
on the two-photon
Hong-Ou-Mandel-type interference 
which is insensitive to the relative phase between two photons.
Several experiments have proven that the path length fluctuations
only need to be kept on the scale of the photon's coherent length,
from hundreds of micrometer \cite{YangPRL2006} to tens of meters
\cite{FelintoNaturePhysics2006,ChanelierePRL2007,ZhenshengPRL2007}.
In our original protocol [10,11], two laser beams with fixed
relative phase are needed to excite two atomic ensembles in order
to generate the atom-photon entanglement for the local
communication node. Only the path length between two ensembles in
the local node need to be stabilized to sub-wavelength scale. Some
recent works close to the requirements of our protocol have
provided the techniques to generate atom-photon entanglement with
spin excitation of magnetic sublevels [16, 17] or dual-species
atomic ensemble to prevent for the propagating phase difference
[18]. But for each of these there remain problems like balancing
the excitation between the ensembles or the complexity and
efficiency of frequency mixing, which make it hard to implement
the full protocol over long distance.

In this Letter, we present a new approach to effectively generate
the entanglement between the atomic qubit and photonic qubit based
on atomic ensemble in a local magneto-optical trap (MOT). This
atom-photon entanglement can serve as a segment of the improved
protocol \cite{zhaoboPRL2007}. Contrast to the previous
experiments
\cite{MatsukevichScience2004,ChouNature2005,MatsukevichPRL2005,RiedmattenPRL2006,LanPRL2007},
only one atomic ensemble is used to be excited by only one write
beam with single frequency. Two spontaneous Raman scattered fields
(anti-Stokes fields) in different spatial modes are combined on a
polarizing beam splitter and serve as the photonic qubit. The
collective spin excitations in the atomic ensemble corresponding
to the two anti-Stokes fields represent the atomic qubit. This new
scheme makes the local phase stabilization simple. With a single
write beam excitation, only the phase difference between the two
selected modes is relevant and can easily be stabilized by the
local build-in Mach-Zehnder interferometer
\cite{ChenYuaoquantph2007}. Besides, high dimensional entanglement
and hyper-entangled state can be easily realized by extend the
approach to select more spatial mode of the collective excitation.

\begin{figure}
\includegraphics[width=8cm]{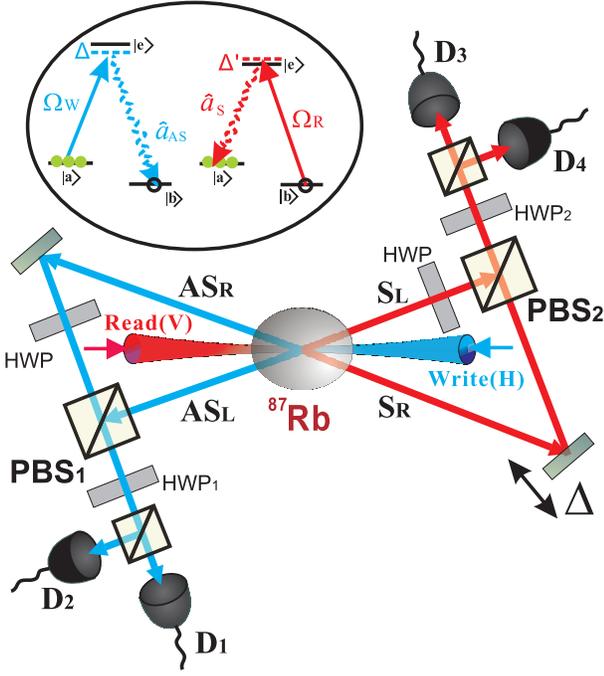}
\caption{Illustration of the scheme of the experiment setup and
the relevant energy levels of the $^{87}$Rb atoms. Cold $^{87}$Rb
atoms captured by MOT are initially prepared in state $|a\rangle$.
A weak write pulse $\Omega_{\mbox{\tiny W}}$ with a beam waist of
240 $\mu$m illuminates the atom cloud to generate the spin
excitation. The spontaneous Raman scattered anti-Stokes field
$AS_L$ and $AS_R$ are detected at $\pm3^{\circ}$ to the
propagating direction of the write beam, with the beam waist of 70
$\mu$m, defining the spatial mode of the atomic ensembles $L$ and
$R$, respectively. The two anti-Stokes field are combined on a
polarizing beam splitter $\mathrm{PBS}_1$ and sent to the
polarization analyzer. This creates the entanglement between the
polarization of the anti-Stokes field and the spatial modes of
spin excitation of atoms in atomic ensemble. After a controlled
storage time $\tau$, the entanglement is verified by retrieving
the spin excitation back to the Stokes fields $S_L$ and $S_R$ by a
strong read pulse, which is overlapped and counter-propagates to
the write beam. After overlap the Stokes fields on
$\mathrm{PBS}_2$, the entanglement can be proved.}
\label{fig:setup}
\end{figure}

The basic setup of our experiment is shown in Fig.\ref{fig:setup}.
A cold $^{87}$Rb atomic cloud with temperature about 100 $\mu$K in
the MOT is used as the medium to generate and store the
information of the quantum excitation. The two hyperfine ground
states $|5S_{1/2},F=2\rangle$=$|a\rangle$ and
$|5S_{1/2},F=1\rangle$=$|b\rangle$ and the excited state
$|5P_{1/2},F=2\rangle$=$|e\rangle$ form a $\Lambda$-type system.
After loading the MOT, the atoms are first pumped to initial state
$|a\rangle$. A single weak 75 ns write beam illuminates the atom
cloud with beam waist of 240 $\mu$m and 10 MHz red-detuned to
$|a\rangle\rightarrow|e\rangle$ transition. Two anti-Stokes fields
(70 $\mu$m waist, $|e\rangle\rightarrow|b\rangle$) $AS_L$ and
$AS_R$ induced by the write beam via spontaneous Raman scattering
are collected at $\pm 3^{\circ}$ relative to the propagating
direction of the write beam. This also defines the spatial mode of
the atomic ensemble $L$ and $R$. With small excitation
probability, the atom-light field can be expressed as
\cite{DuanNature2001}
\begin{eqnarray}\label{eqn:excitation}
|\Psi\rangle_{m}\sim|0_{\mbox{\ssmall AS}}0_b\rangle_{m}+
\sqrt{\chi_m}|1_{\mbox{\ssmall AS}}1_b\rangle_{m}+O(\chi_m),
\end{eqnarray}
where $\chi_m\ll1$ is the excitation probability of one collective
spin in ensemble $m$ ($m=L,R$), and
$\sqrt{\chi_m}|i_{\mbox{\ssmall AS}}i_b\rangle_{m}$ denote the
$i$-fold excitation of the anti-Stokes light field and the
collective spin in atomic ensemble.

When the write beam excites the atomic ensemble and an anti-Stokes
photon is generated, it also transfers the momentum to the
collective spin excitation in the atomic ensemble. To fulfill the
momentum conservation, the overall $k$-vector of the collective
excitation after the spontaneous Raman scattering is
${\vec{k}_{atom}}=\vec{k}_{W}-\vec{k}_{AS}$, where $\vec{k}_{AS}$
and $\vec{k}_{W}$ are the wave vector of the anti-Stokes field and
write beam, respectively. If no other external field interrupts
the atomic state, during the storage time $\tau$, the momentum of
the collective excitation is kept. When the read pulse is applied
on the atomic ensemble to retrieve the collective excitation back
into a correlated Stokes field, the momentum of the collective
excitation is transfered back to the Stokes field. The wave vector
of the Stokes field becomes
$\vec{k}_{S}=\vec{k}_{R}+\vec{k}_{atom}$, where $\vec{k}_{R}$
represents the wave vector of the read beam. Then after the
retrieve process, the wave vector of the correlated Stokes field
fulfill the mode-matching condition \cite{BrajePRL2004}
\begin{eqnarray}
\vec{k}_{S}=\vec{k}_{R}+\vec{k}_{W}-\vec{k}_{AS}.
\end{eqnarray}
Under the counter-propagating condition of read and write beams
(shown in Fig. \ref{fig:setup}), the anti-Stokes and mode-matched
Stokes fields are also counter-propagating
($\vec{k}_{S}\simeq-\vec{k}_{AS}$).

To characterize the light field, we measure the cross correlation
$g^{(2)}_{AS,S}$ \cite{KuzmichNature2003,ChenShuaiPRL2006}, which
marks the degree of quantum correlation, between the anti-Stokes
and the Stokes fields. As two anti-Stokes fields $AS_L$ and $AS_R$
are detected at two different spatial modes, two corresponding
Stokes fields $S_L$ and $S_R$ can be detected during the retrieve
process. For the mode-matched fields $S_L$ and $AS_L$ ($S_R$ and
$AS_R$), the cross correlation $g^{(2)}_{AS,S}\gg1$ when
$\chi\ll1$, which means good quantum correlation between those
fields. But for the unmatched fields $S_L$ and $AS_R$ ($S_R$ and
$AS_L$), no quantum correlation is observed
($g^{(2)}_{AS,S}\sim1$), which means there's no cross talk between
these two different modes. The viability of our new approach is
guaranteed by this condition.

For the further part of our experiments, we adjust the two modes
$L$ and $R$ to be equal ($\chi_L=\chi_R=\chi$), select orthogonal
polarization of the two anti-Stokes fields, combine them on a beam
polarizing beam splitter $\mathrm{PBS}_1$ and send into a
polarization analyzer, as illustrated in Fig. \ref{fig:setup}.
Neglecting the vacuum state and high order excitations, the
entangled states between the photonic and the atomic qubit can be
described as,
\begin{eqnarray}\label{eqn:entanglement_1}
|\Psi\rangle=\frac{1}{\sqrt{2}}(|H\rangle|L\rangle+e^{i\phi_{1}}|V\rangle|R\rangle)
\end{eqnarray}
where $|H\rangle$/$|V\rangle$ denotes horizontal/vertical
polarizations of the single anti-Stokes photon and
$|L\rangle$/$|R\rangle$ denotes single collective spin excitation
in ensemble $L$/$R$, $\phi_{1}$ is the propagating phase
difference between the two anti-Stokes fields before they overlap
at $\mathrm{PBS}_{1}$. Physically, the atom-photon entangled state
(\ref{eqn:entanglement_1}) is equivalent to the maximally
polarization entangled state generated by spontaneous parametric
down-conversion \cite{KwiatPRL1995}.

To verify the entanglement between the anti-Stokes field and the
atomic spin excitation, a relative strong read pulse with 75 ns
close to resonance of $|e\rangle\rightarrow|b\rangle$ transition
counter-propagating with the write beam is applied after a
controllable time $\tau$ to convert the atomic collective
excitation back into Stokes fields.

After combine the two Stokes fields on $\mathrm{PBS}_2$ (see Fig.
\ref{fig:setup}), the superposition state of anti-Stokes and
Stokes fields is the following maximally polarization entangled
state
\begin{eqnarray}
|\Psi\rangle_{AS,S}=\frac{1}{\sqrt{2}}|H\rangle_{AS}|V\rangle_{S}
+ e^{i(\phi_{1}+\phi_{2})}|V\rangle_{AS}|H\rangle_{S},
\end{eqnarray}
where $\phi_{2}$ represent the propagating phase difference
between two Stokes fields before they overlap at the
$\mathrm{PBS}_2$. In our experiment, the total phase
$\phi_{1}+\phi_{2}$ is actively stabilized via the build in
Mach-Zehnder interferometer and fixed to zero
\cite{ChenYuaoquantph2007}.

\begin{figure}
  \includegraphics[width=8cm]{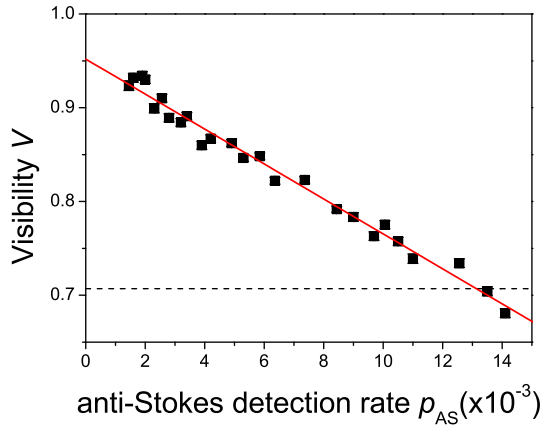}\\
  \caption{Visibility of the interference fringes $V$ between
  Anti-Stokes fields and Stokes fields various with the changing
  of the detected rate of anti-Stokes field $p_{AS}$. The solid
  line is the fit corresponding to Eq.(\ref{eqn:visibility}). The
  dashed line shows the bound of $1/\sqrt{2}$ which mark the
  limit to violate the CHSH-type Bell's inequality.}\label{fig:visibility}
\end{figure}

To investigate the scaling of entanglement with the excitation
probability $\chi$, we measure the visibility $V$ of the
interference fringes of the coincidence rate between anti-Stokes
and Stokes photons for various value of $\chi$ with fixed memory
time $\tau=500$ ns. The half waveplate $\mathrm{HWP}_{1}$ (see
Fig. \ref{fig:setup}) is set to $+22.5^{\circ}$ to measure the
anti-Stokes fields under $|H\rangle+|V\rangle$ base and rotate
$\mathrm{HWP}_{2}$ to measure the Stokes fields under different
bases. As $\chi$ increases, the high order term in Eqn.
(\ref{eqn:excitation}) can not be neglected. The visibility $V$
can be expressed as the function of cross correlation between the
anti-Stokes and Stokes fields \cite{RiedmattenPRL2006}
\begin{eqnarray}
V=\frac{g^{(2)}_{AS,S}-1}{g^{(2)}_{AS,S}+1}.
\end{eqnarray}
Ideally, the relationship of the excitation rate $\chi$ and cross
correlation $g^{(2)}_{AS,S}$ is $g^{(2)}_{AS,S}=1+1/\chi$.
Considering the total detected efficiency of the anti-Stokes field
$\eta_{AS}$, we have the detection rate of the anti-Stokes photon
$p_{AS}=\eta_{AS}\chi$, where $\eta_{AS}$ is the detect efficiency
of the anti-Stokes channel. At the small excitation rate limit
($\chi\ll1$), the visibility can be expressed as
\begin{eqnarray}\label{eqn:visibility}
V=1-2p_{AS}/\eta_{AS}.
\end{eqnarray}
In our experiment, $\eta_{AS}\sim8\%$. Figure \ref{fig:visibility}
shows the measured visibility $V$ varing with $p_{AS}$. As the
intensity of the write beam is tuned to make the excitation rate
$\chi$ decrease, which corresponds to decrease of $p_{AS}$, the
visibility $V$ increases as does the degree of entanglement. The
solid line is the linear fit for the experiment data. At
$p_{AS}\rightarrow0$, $V$ is near 0.95. This imperfection is
mainly caused by the overlap of the two anti-Stokes fields $AS_L$
and $AS_R$, the noise of the detector and the phase fluctuation in
the interferometer. As the detection rate $p_{AS}$ increases, the
probability of high order excitations increases faster than that
of the single excitation. Then the correlation $g^{(2)}_{AS}$
decreases, as well as the visibility. At
$p_{AS}<1.3\times10^{-2}$, $V$ is larger than $1/\sqrt{2}$ which
marks the bound of violation of the Clauser-Horne-Shimony-Holt
(CHSH) type Bell's inequality
\cite{RiedmattenPRL2006,ClauserPRL1969}.

\begin{figure}
  \includegraphics[width=8cm]{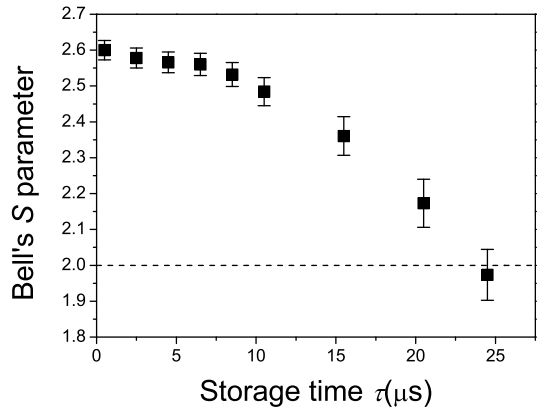}\\
  \caption{Decay of the $S$ parameter in the Bell's inequality
  measurement with the storage time $\tau$. The dashed line
  shows the classical bound of $S=2$.}\label{fig:BellInequality}
\end{figure}

To further study the storage ability of the atomic ensemble
quantum memory, we characterize the temporal decay of entanglement
with storage time $\tau$. Here we measure the decay of the $S$
parameter, sum of the correlation function in CHSH type Bell's
inequality, where $S\leq2$ for any local realistic theory with
\begin{eqnarray}\label{eqn:chsh}
S=|E(\theta_1,\theta_2)-E(\theta_1,\theta_2^\prime)-
E(\theta_1^\prime,\theta_2)- E(\theta_1^\prime,\theta_2^\prime)|,
\end{eqnarray}
Here $E(\theta_1,\theta_2)$ is the correlation function, where
$\theta_1$ and $\theta_2^\prime$ ($\theta_1^\prime$ and
$\theta_2^\prime$) are the measured polarization bases of the
anti-Stokes field and Stokes field. During the measurement, the
$\mathrm{HWP_{1}}$ and $\mathrm{HWP_{2}}$ are set to different
angles to make the bases settings at ($0^\circ$,$22.5^\circ$),
($0^\circ$,$-22.5^\circ$), ($45^\circ$,$22.5^\circ$) and
($45^\circ$,$-22.5^\circ$), respectively. The excitation rate
$\chi$ was fixed to get $p_{AS}=2\times10^{-3}$, and the result of
measurement is shown in Fig. \ref{fig:BellInequality}. At the
storage time of 500 ns, $S=2.60\pm0.03$, which violates Bell's
inequality by 20 standard deviations. As the storage time
increases, the $S$ parameter decreases, indicating the decoherence
of the entanglement. At storage time $\tau=$20.5 $\mathrm{\mu s}$,
we still get $S=2.17\pm0.07$, which means the character of quantum
entanglement is still well preserved. The decay of $S$ parameter
with increasing storage time $\tau$ is caused by the residual
magnetic field which inhomogeneously broadens the ground state
magnetic sublevels. This process can be observed from the decay of
the retrieve efficiency and the cross correlation between
anti-Stokes and Stokes fields.

\begin{figure}
\includegraphics[width=8cm]{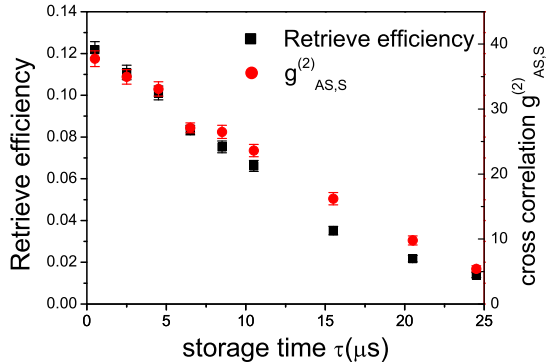}
\caption{The decay of retrieve efficiency and cross correlation
$g^{(2)}_{12}$ with the storage time $\tau$. The anti-Stokes
detection rate is fixed at $p_{AS}=2\times10^{-3}$. The square
dots show the decay process of the retrieve efficiency of the
Stokes fields, round dots show the decay of the cross correlation
$g^{(2)}_{AS,S}$ between anti-Stokes field and Stokes field.
}\label{fig:retieveg12}
\end{figure}

Shown in Fig. \ref{fig:retieveg12}, the retrieve efficiency and
the cross correlation between anti-Stokes and Stokes field all
decreases with increasing the storage time $\tau$. At $\tau=500$
ns, the overall retrieve efficiency (including the transmission
loss and the detector efficiency) is $12.2\pm0.4\%$ and the cross
correlation $g^{(2)}_{AS,S}=38\pm1$. At $\tau=$20.5 $\mu$s, the
retrieve efficiency and cross correlation decrease to
$2.2\pm0.1\%$ and $g^{(2)}_{AS,S}=9.8\pm0.7$, respectively. These
values are still sufficient to violate the CHSH-type Bell's
inequality. When $\tau$ is longer than $24\mu s$,
$g^{(2)}_{AS,S}<6$ makes it insufficient to violate the Bell's
inequality.

In conclusion, we have generated a robust atom-photon entanglement
with a novel approach. A single write beam and a single atomic
ensemble are used to generate the collective spin excitations. Two
spatial modes of collective excitations are defined by the
collection modes of anti-Stokes fields. The conservation of
momentum during the atom-photon interaction prevent for the cross
talk between different excited spatial modes. The visibility of
the entanglement and violation of the CHSH type Bell's inequality
are measured to prove the atom-photon entanglement between
anti-Stokes photon and collective excitation in atomic ensemble.
Also with the help of the build-in quantum memory, the violation
of the Bell's inequality still exists after 20.5 $\mu$s,
corresponding to the time of light propagating 4 km in an optical
fiber. That means we have successfully achieved a memory build-in
atom-photon entanglement source which can work as a node of the
long-distance quantum communication networks. Further more, if the
atomic ensemble is confined in the optical trap and "clock states"
\cite{HarberPRA2002} is implemented, the memory time could be
extended to longer than 1 ms. Moreover, if more anti-Stokes modes
are selected at different angles corresponding to the write beam,
this approach can be easily extended to generate high order
entanglement, which could be very useful in the complex quantum
cryptography and quantum computation.

This work was supported by the Deutsche Forschungsgemeinschaft
(DFG), the Alexander von Humboldt Foundation, the Marie Curie
Excellence Grant of the EU, the Deutsche Telekom Stiftung and the
CAS.


\begin{thebibliography}{99}

\bibitem{BriegelPRL1998}
H.-J. Briegel, W.~D\"{u}r, J.~I. Cirac, and P.~Zoller, Phys. Rev. Lett.
  \textbf{81}, 5932 (1998).


\bibitem{DuanNature2001}
L.-M. Duan, M.~D. Lukin, J.~I. Cirac, and P.~Zoller, Nature (London)
  \textbf{414}, 413 (2001).

\bibitem{KuzmichNature2003}
A. Kuzmich \emph{et~al.}, Nature (London) \textbf{423}, 731
  (2003).



\bibitem{ChaneliereNature2005}
T. Chaneli\`{e}re \emph{et~al.}, Nature (London) \textbf{438}, 833
(2005).

\bibitem{EisamanNature2005}
M. D. Eisaman \emph{et~al.}, Nature (London) \textbf{438}, 837
(2005).

\bibitem{MatsukevichScience2004}
D. N. Matsukevich and A. Kuzmich, Science \textbf{306}, 663
(2004).

\bibitem{ChouNature2005}
C.~W. Chou \emph{et~al.}, Nature (London) \textbf{438}, 828
(2005).

\bibitem{ChenShuaiPRL2006}
S.~Chen \emph{et~al.}, Phys. Rev. Lett. \textbf{97}, 173004
(2006).

\bibitem{ChouScienceExpress2007}
C. W. Chou \emph{et~al.}, Science Express,
10.1126/science.1140300.

\bibitem{zhaoboPRL2007}
B.~Zhao \emph{et~al.}, Phys. Rev. Lett. \textbf{98}, 240502
(2007).

\bibitem{ChenZBquantph2006}
Z. B. Chen \emph{et~al.}, arXiv:quant-ph/0609151v1, to appear in
Phy. Rev. A (2007).


\bibitem{YangPRL2006}
T.~Yang \emph{et~al.}, Phys. Rev. Lett. \textbf{96}, 110501
(2006).

\bibitem{FelintoNaturePhysics2006}
D. Felinto \emph{et~al.}, Nature Physics \textbf{2}, 844 (2006).

\bibitem{ChanelierePRL2007}
T. Chaneli\`{e}re \emph{et~al.}, Phys. Rev. Lett. \textbf{98},
113602 (2007).

\bibitem{ZhenshengPRL2007}
Z. -S. Yuan \emph{et~al.}, Phys. Rev. Lett. \textbf{98}, 180503,
(2007).

\bibitem{MatsukevichPRL2005}
D. N. Matsukevich \emph{et~al.}, Phys. Rev. Lett. \textbf{95},
040405 (2005).

\bibitem{RiedmattenPRL2006}
H.~de~Riedmatten \emph{et~al.}, Phys. Rev. Lett. \textbf{97},
113603 (2006).

\bibitem{LanPRL2007}
S.-Y.Lan \emph{et~al.} Phys. Rev. Lett. \textbf{98}, 123602
(2007).

\bibitem{ChenYuaoquantph2007}
Y.-A. Chen \emph{et~al.}, arXiv:0705.1256 (2007).

\bibitem{BrajePRL2004}
D.~A.~Braje \emph{et~al.}, Phys. Rev. Lett. \textbf{93}, 183601
(2004).

\bibitem{ClauserPRL1969}
J.~F. Clauser, M.~Horne, A.~Shimony, and R.~A. Holt, Phys. Rev.
Lett. \textbf{23}, 880 (1969).

\bibitem{KwiatPRL1995}
P. Kwiat \emph{et~al.}, Phys. Rev. Lett. \textbf{75}, 4337 (1995).

\bibitem{HarberPRA2002}
D. M. Harber \emph{et~al.}, Phys. Rev. A \textbf{66}, 053616
(2002).









\end{thebibliography}
\end{document}